VERSION 2017-04-25a

**Reply to Hicks *et alii* (2017) Reply to Morrison *et alii* (2016) Refining the relevant population in forensic voice comparison – A response to Hicks *et alii* (2015) The importance of distinguishing information from evidence/observations when formulating propositions**

Dear Editor:

We find that Hicks *et aliorum* reply [1] to our response [2] to their original paper [3] mostly resolves apparent disagreements between us. We are in agreement that forensic practitioners must think carefully about, carefully define, and carefully communicate to the court, the propositions that they address. We further agree that forensic practitioners should take care to avoid the sorts of errors described in [3].

It is very helpful that [1] emphasizes that [3] was intended to present a "framework for thinking", not a "prescriptive" "list of default propositions". Our response in [2] was motivated by the fear that the section in [1] on forensic voice comparison would be read prescriptively, and that this would lead to negative outcomes.

An apparent point of disagreement outstanding is whether properties such as speaker sex, language spoken, and accent spoken should be treated as "evidence", or as "information" which can be used to delimit the relevant population. This depends on the circumstances of the case. For example, if the sex of the speaker is unclear or in dispute, then the relevant population cannot be delimited to speakers of a particular sex. In the section from [2] quoted in [1], we were careful to state that speaker sex, language spoken, and accent spoken will *usually* be obvious, and will *usually* be perceptually salient to all parties, and will therefore *usually* not be disputed. *In the latter circumstances*, the court would be unlikely to consider expert testimony regarding these properties (e.g., whether the speaker is male of female) admissible, since it would not require "scientific, technical, or other specialized knowledge" (US Federal Rule of Evidence 702), nor would it "provide the court with information likely to be outside the court's own knowledge and experience" (England & Wales Criminal Practice Directions 19A.1). Thus, *in such circumstances*, it would not be advisable for a forensic practitioner to attempt to provide expert testimony regarding these properties, e.g., whether the speaker is male or female.[1] Instead, we continue to argue that it would be appropriate for the forensic practitioner to use these properties to delimit the relevant population, and explain to the court that this is what they have done. And in fact Hicks *et alii* appear to agree with us on this. Referring to an example based on regional accent, they state that "whether or not the observations on the accent are contested, ... one can disclose this element and say that the evaluation is based on the assumption that the relevant population is a group of people from a given region." It seems that Hicks *et alii* do not actually disapprove of the procedures that we describe in [2] for conducting a forensic voice comparison analysis. Their stated disagreement appears to be related to definitions of words, rather any real disagreement regarding principles.

---

[1] Note that this is a different question from that of what strength of evidence may be associated with the fact that both the known- and questioned speaker are of a given sex, or speakers of a given language or accent as was our example in [2].

In [2] we explicitly stated that "We do not mean to imply that there is something special about forensic voice comparison which makes it different from all other branches of forensic science." This was intended to prevent a misinterpretation of our remarks as implying that forensic voice comparison was somehow exempt from principles that are generally applicable in evaluation of strength of evidence across all branches of forensic science, or that only those with casework experience in forensic voice comparison are competent to opine on the application of those principles within this branch of forensic science. Unfortunately, Hicks *et alii* appear to have misinterpreted our remarks in just this way.

The history of forensic voice comparison includes practitioners of the spectrographic approach arguing that only those with experience using that approach should be allowed to testify which respect to its validity. The approach was widely criticised by others on the grounds that its validity had not been empirically demonstrated under casework conditions. One does not need to have experience using a particular approach to be competent to argue the principle that, whatever the approach, it should not be used unless it has been empirically validated under conditions sufficiently similar to the conditions of intended use. We hereby explicitly state that it is *not* our opinion that only those with first-hand experience in a particular branch of forensic science or with a particular approach are competent to opine on the application of principles to that branch of forensic science or to that approach.

A point that we wanted to make in [2] is, perhaps, more subtle. The principles are absolutely applicable, but, without sufficient understanding of the complexities of a particular branch of forensic science, one may inadvertently misapply the principles, leading to undesired results. In general, unless authors are themselves experts in a particular branch of forensic science, we would recommend that they collaborate or at least consult with experts in that branch of forensic science before basing arguments on examples drawn from that branch of forensic science. This would hopefully reduce the potential for misunderstanding and miscommunication, and more effectively and efficiently achieve shared goals.

Sincerely

*Geoffrey Stewart Morrison* \*

    Independent Forensic Consultant, Vancouver, British Columbia, Canada

    Department of Linguistics, University of Alberta, Edmonton, Alberta, Canada

*Ewald Enzinger*

    Eduworks, Corvallis, Oregon, United States of America

*Cuiling Zhang*

    School of Criminal Investigation, Southwest University of Political Science and Law, Chongqing, China

    Chongqing Institutes of Higher Education Key Forensic Science Laboratory, Chongqing, China

* Corresponding author. *E-mail address*: geoff-morrison@forensic-evaluation.net

**Editor's response**

While this is a good response, we've already had 'their paper → your response → their response to your response', and this format is the norm in these matters otherwise we could go on forever.